\documentclass[%
aps,prl,reprint,twocolumn,nofootinbib,floatfix,superscriptaddress ]{revtex4-2}

\usepackage{graphicx}
\usepackage{dcolumn}
\usepackage{bm}

\usepackage[english]{babel}

\usepackage{multirow}
\usepackage{amssymb}
\usepackage{amsmath,bbm}
\usepackage[figuresright]{rotating}
\usepackage{capt-of}

\protected\def\lc{C}

\newcommand{\La}{{\Lambda}}
\newcommand{\Si}{{\Sigma}}

\newcommand{\be}{\begin{eqnarray}}
\newcommand{\ee}{\end{eqnarray}}

\newlength{\feynwidth} 
\newlength{\feynwidthbig} 

\usepackage{xcolor} 

\usepackage{hyperref}
\usepackage{graphicx}

\hypersetup{colorlinks=true,citecolor= blue}

\usepackage[english]{babel}
\usepackage[autolanguage]{numprint}

\begin{document}

\title{Light $\Lambda$ hypernuclei studied with chiral 
hyperon-nucleon\\ and hyperon-nucleon-nucleon forces}
\author{Hoai Le}%
\email{h.le@fz-juelich.de}
\affiliation{Institute for Advanced Simulation, 
Forschungszentrum J\"ulich, D-52425 J\"ulich, Germany} 

\author{Johann Haidenbauer }%
\email{j.haidenbauer@fz-juelich.de}
\affiliation{Institute for Advanced Simulation, 
Forschungszentrum J\"ulich, D-52425 J\"ulich, Germany}

\author{Ulf-G. Mei{\ss}ner }%
\email{meissner@hiskp.uni-bonn.de}
\affiliation{Helmholtz-Institut~f\"{u}r~Strahlen-~und~Kernphysik~and~Bethe~Center~for~Theoretical~Physics,
~Universit\"{a}t~Bonn,~D-53115~Bonn,~Germany}
\affiliation{Institute for Advanced Simulation, 
Forschungszentrum J\"ulich, D-52425 J\"ulich, Germany}
\affiliation{Tbilisi State University, 0186 Tbilisi, Georgia}
\affiliation{CASA, Forschungszentrum J\"ulich, D-52425 J\"ulich, Germany}

\author{Andreas Nogga}
\email{a.nogga@fz-juelich.de}
\affiliation{Institute for Advanced Simulation, 
Forschungszentrum J\"ulich, D-52425 J\"ulich, Germany}
\affiliation{CASA, Forschungszentrum J\"ulich, D-52425 J\"ulich, Germany}

\date{\today}

\begin{abstract}
A study of light $\Lambda$ hypernuclei in chiral 
effective field theory is presented. For the first
time chiral $\Lambda$NN and $\Sigma$NN three-body
forces are included consistently. The calculations
are performed within the no-core shell model. 
Results for the separation energies of the
hypernuclei 
$^3_{\Lambda}\mathrm{H}$, 
$^4_{\Lambda}\mathrm{H}$/$^4_{\Lambda}\mathrm{He}$,
$^5_{\Lambda}\mathrm{He}$, and $^7_{\Lambda}\mathrm{Li}$ are given. 
It is found that the experimental values can be
fairly well reproduced once YNN three-body forces are
taken into account. 
\end{abstract}


\maketitle

{\bf Introduction.}--
Three-body forces (3BFs) have a long and varied history in nuclear physics~\cite{Kalantar:2011rzs}.
Initially, they were introduced on a phenomenological basis, in order to explain the 
observed discrepancy between the measured
binding energies of light nuclei such as the triton
or $^4$He and the predictions based on realistic 
nucleon-nucleon (NN) potentials.
The concept of 3BFs changed radically in the early 1990s with
the pivotal works of Weinberg \cite{Weinberg:1990bf,Weinberg:1992jd}. He introduced a framework that allows one to construct NN and 3N forces consistently, namely chiral effective field theory (EFT), together with a pertinent power
counting which establishes a hierarchy for the 
relative strength/importance
of 3N (4N, ...) forces in relation to that of the 
NN interaction. 
Performing few-nucleon calculations  based on this scheme
is by now standard
\cite{Epelbaum:2008ga,Navratil:2007we,Maris:2020qne,Otsuka:2009cs,Hu:2021trw,Elhatisari:2022zrb}. 

The 3BFs included in early hypernuclear physics calculations were likewise of purely 
phenomenological nature \cite{Bhaduri:1967,Gal:1971gb,Bodmer:1984gc,Usmani:1995fma}. However,
unlike the present situation in the
few-nucleon sector mentioned above, even now
hyperon-nucleon-nucleon 
(YNN) forces are still considered on a phenomenological 
level in essentially all pertinent investigations.
This concerns not only studies of ordinary
hypernuclei \cite{Lonardoni:2013rm,Lonardoni:2017uuu},
but in particular the discussion on the role that
hyperons play for the size and stability of neutron stars 
\cite{Chatterjee:2015pua,Schaffner-Bielich:2020psc,Tolos:2020aln}.
Specifically, for the latter issue,
YNN 3BFs are seen as one of the most
promising solutions for the so-called hyperon
puzzle, i.e. the quest to reconcile the 
appearance of $\Lambda$'s (and possibly other
hyperons) at densities realized in compact objects like neutron stars, 
causing a softening of the equation of state (EOS),
with the observed mass-radius relation that can
be explained only with a stiff EOS
\cite{Vidana:2010ip,Lonardoni:2014bwa,Gerstung:2020ktv,Wirth:2016iwn}. 
Given their key role for this topic, it is vitally important to put the YNN 3BFs on a more solid
ground.

Indeed, the first 
derivation of 3BFs involving baryons with strangeness ($\Lambda$, $\Sigma$, $\Xi$), 
based on SU(3) chiral EFT, was  given only
a few years ago~\cite{Petschauer:2016ho}. In addition, recently
the $\Lambda$N-$\Sigma$N interaction has been established up 
to next-to-next-to-leading order (N$^2$LO) in the chiral
expansion \cite{Haidenbauer:2023qhf}, i.e.
the order at which the leading YNN force formally appears
in the Weinberg counting. Thus all ingredients
for a consistent inclusion of hyperon-nucleon (YN) and YNN forces 
in calculations of $\Lambda$ hypernuclei are 
 available now. Fortunately, on the computational
side, so-called {\it ab initio} methods like the 
no-core shell model (NCSM) have reached a level where rigorous calculations of hypernuclei
up to A=8 \cite{Le:2022ikc} 
(or even up to A=13 \cite{Wirth:2014apa}) 
based on state-of-the-art YN and YNN potentials
have become feasible.

In this letter, we present the
first calculation of light $\La$ hypernuclei (A=3-7) employing YN 
and YNN interactions, both derived from chiral EFT. 
The essential accomplishment of this work is a consistent treatment of the YN and YNN forces up to N$^2$LO in the chiral expansion. To the best of our knowledge,
there is only one other calculation of hypernuclei (A=3-5) with similar ambition, 
which includes consistently $\La$N and $\La$NN
forces~\cite{Contessi:2018qnz}. However, it is performed 
in pionless EFT at leading order (LO). 
A recent study of hypernuclei within nuclear lattice
effective field theory is likewise restricted to
LO $\Lambda$N forces~\cite{Hildenbrand:2024ypw} and similarly for hyper-neutron matter~\cite{Tong:2024egi}.
There are already calculations that include the chiral 
YNN force together with chiral YN interactions
\cite{Kamada:2023txx,Kohno:2023xvh}. 
However, those have only exploratory character
because the strength of the 3BF, encoded in so-called
low-energy constants (LECs), was fixed using dimensional arguments, and not 
by considering actual
separation energies. Moreover, these studies are restricted to the hypertriton. 
In the present work we determine the strength of the 3BF
by adjusting the LECs to the 
$^4_{\Lambda}\mathrm{H}$/$^4_{\Lambda}\mathrm{He}$ and
$^5_{\Lambda}\mathrm{He}$ bound states. 
Thereby we want to answer two important questions:
(i) Can the spectrum of A=3-7 hypernuclei be described by including 3BFs, and, if so,  how well? 
(ii) Are the resulting 3BFs consistent with 
the size expected from chiral power counting?


{\bf No-core shell model.}--
We apply the Jacobi NCSM 
for calculating the $\Lambda$-separation 
(binding) energies of  the A=3-7 hypernuclei. A detailed description of the formalism and of the procedure to extract the binding (separation) energies can be found in Ref.~\cite{Le:2020zdu}.  
In all calculations, contributions 
   of the NN(YN) potentials in partial waves up to $J = 6(5)$ 
   are included, while for the 3N interaction all partial waves with total angular momentum $ J_{\mathrm{3N}} \leq 9/2$ are taken into account. It has been checked that higher partial waves only contribute negligibly compared to the 
   harmonic oscillator (HO) model space uncertainties.  
   In order to speed up the convergence of the NCSM with respect to the model space, all the employed  NN, 3N and YN potentials are 
   SRG-evolved to a flow parameter of $\lambda=1.88$~fm\textsuperscript{-1},  see \cite{Le:2020zdu} and references therein.    The latter is commonly  used in nuclear calculations, which, on the one hand, yields rather well-converged nuclear binding energies, and on the other hand, minimizes the possible contribution of SRG-induced 4N and higher-body forces \cite{LENPIC:2022cyu}. 
   Furthermore, in all the calculations, the SRG-induced 
   YNN interaction with the total angular momentum $J_{\mathrm{YNN}} \leq 5/2$ is included. Based on the contributions of $J_{\mathrm{YNN}} \leq 1/2$, $3/2$ and $5/2$, the contribution from  higher partial waves  $J_{\mathrm{YNN}} \ge 7/2$ is estimated to be negligibly  small and, therefore, is omitted from the calculations.
   With the proper inclusion of these SRG-induced three-body forces, the 
   otherwise strong dependence of the $\Lambda$ separation energies
   on the SRG-flow parameter \cite{Wirth:2014ko,Le:2020zdu} is  practically removed for the $A=3,4$ systems and remains
   insignificantly small for $^5_{\Lambda}\mathrm{He}$ \cite{Wirth:2018ho,Le:2023bfj}.
   All calculations in this letter will be based on the SMS
   NN potential 
   N\textsuperscript{4}LO\textsuperscript{+}
   \cite{Reinert:2017usi} with a cutoff of $\Lambda =550$~MeV and include a 
   corresponding N$^2$LO 3NF \cite{Le:2023bfj}. For the YN interaction the SMS potential 
   $\mathrm{N^2LO}$(550$^b$) from 
   Ref.~\cite{Haidenbauer:2023qhf} 
   (also with $\Lambda =550$~MeV) is employed.
   In addition, for the first time, the full leading 
   chiral YNN forces at N\textsuperscript{2}LO
   are taken into account. First exploratory studies of the contributions of the chiral $\Lambda \mathrm{N N}$ forces to the separation energies have been performed for the $A=3$ hypernucleus \cite{Kamada:2023txx,Kohno:2023xvh} and  $A=3-5$ systems \cite{Le:2024} using a non-local regulator.


{\bf Structure of the three-baryon force.}--
The general structure of the leading 3BFs within 
SU(3) chiral EFT
has been worked out in detail in
\cite{Petschauer:2016ho}.
Like in the 3N case \cite{Epelbaum:2002vt,Epelbaum:2008ga}, 
there are contributions to the 3BF from two-meson
exchanges,  one-meson exchanges and  6-baryon
contact terms. However, since for YNN the
Pauli principle is less restrictive, 
the corresponding 3BFs have a much richer
structure. For example, regarding the contact term,
there is only one such contribution
to the 3N force, but eight in the $\Lambda$NN$-$$\Sigma$NN
system \cite{Petschauer:2016ho}.  
To be concrete, while the three-nucleon contact 
potential in its antisymmetrized form (using the antisymmetrizer ${\cal A}$) is given by 
\cite{Epelbaum:2002vt} 
\begin{equation}
V^{\rm 3N}_\mathrm{ct} = \frac12 E\, \mathcal A\, \sum_{j\neq k}\vec\tau_j\cdot\vec\tau_k \,,
\end{equation} 
the corresponding expression for the $\Lambda \mathrm{NN} \rightarrow \Lambda \mathrm{NN}$ potential from the contact term is \cite{Petschauer:2016ho},
 \begin{eqnarray}\label{eq:LNNct}
V^{\Lambda {\rm NN}}_\mathrm{ct} ={}
& \phantom{{},{}} \lc'_1\ (\mathbbm1 - \vec\sigma_2\cdot\vec\sigma_3 ) ( 3 + \vec\tau_2\cdot\vec\tau_3 ) \nonumber\\
& + \lc'_2\ \vec\sigma_1\cdot(\vec\sigma_2+\vec\sigma_3)\,(\mathbbm1 - \vec\tau_2\cdot\vec\tau_3) \nonumber \\
& + \lc'_3\ (3 + \vec\sigma_2\cdot\vec\sigma_3 ) ( \mathbbm1 - \vec\tau_2\cdot\vec\tau_3 ) \,.
\end{eqnarray}
Here, the $\vec\sigma_i$ and $\vec\tau_i$ are the standard
Pauli operators for the spin of the baryons and the isospin of the nucleons, and $E$
and the $C'$'s are low-energy constants (LECs) that
parameterize the strength of the contact interactions.
The situation is similar for the one-meson 
exchange 3BF. In fact, considering only the contribution
from one-pion exchange alone, there is one term (LEC) for 3N, two terms (LECs) for $\La$NN \cite{Petschauer:2016ho}, 
and already six further terms with the 
$\La$NN-$\Si$NN transition potential.

Since the experimental
uncertainty of the $^3_\Lambda$H separation 
energy \cite{HypernuclearDataBase}
is comparable to the expected contribution from the 
$\Lambda$NN 3BF \cite{Le:2023bfj,Le:2024},
it is not practical to determine the pertinent LECs from the hypertriton. 
Therefore, as argued in Ref.~\cite{Haidenbauer:2023qhf}, the only possible and viable way to fix
the 3BFs is via studies of the 
$^4_\Lambda$H/$^4_\Lambda$He
and $^5_\Lambda$He systems. This strategy is followed
here. Still, as should be clear from
the discussion above,
the number of LECs for the leading YNN 3BF exceeds by far the
available experimental constraints from light hypernuclei. Even if we consider only the
$\Lambda$NN 3BF we would have to deal with
5 LECs (3 for the contact terms and 2 from 
the one-pion exchange 3BF) to be compared with only
2 LECs in the 3N case \cite{Epelbaum:2002vt}. 

In the present paper, we restrict
ourselves to the pion-exchange contributions, in line with the 
SMS YN potential \cite{Haidenbauer:2023qhf} where two-meson 
contributions involving the $K$ and/or $\eta$ were neglected
(which is also consistent with large-N$_C$ arguments~\cite{Dashen:1994qi,VM}), 
and 
we exploit the mechanism of resonance saturation
via decuplet baryons ($\Delta$, $\Sigma^*$, $\Xi^*$) to 
estimate the strengths of 
chiral 3BFs 
\cite{Petschauer1:2017gd,Petschauer:2016tee,Petschauer:2020urh}.
Indeed, by
involving decuplet baryons (B$^*$) the resulting 
two-pion exchange $\La$NN and $\Si$NN 3BFs are completely fixed by the 
$\Delta\to N\pi$ decay
width and SU(3) flavor symmetry. The number of independent LECs for the one-pion exchange 3BFs 
and the contact term is strongly reduced, namely 
to two for the $\La$NN$-$$\Si$NN 3BFs and 
to a single LEC for $\La$NN. For example, the 
LECs of the $\La$NN contact interaction in 
Eq.~(\ref{eq:LNNct}) can then be written in terms of
a simple combination of the two LECs of the 
leading order $BB\to BB^*$ 
contact interaction $H_1$ and $H_2$ in SU(3)
chiral EFT, see
Refs.~\cite{Petschauer1:2017gd,Petschauer:2016tee},
and amount to 
\begin{equation}
C'_1 = C'_3 = \frac{(H_1 + 3H_2)^2}{72\Delta}, \quad
C'_2=0 , 
\label{eq:Decu}
\end{equation}
where $\Delta$ is the average mass splitting 
between the octet and decuplet baryons, 
$\Delta\approx 300$~MeV. 
The contact interaction for $\La$NN-$\Si$NN and 
$\Si$NN involve other combinations of $H_1$ and $H_2$
\cite{Petschauer1:2017gd,Petschauer:2016tee}. 

\begin{table*}[t]
\renewcommand{\arraystretch}{1.4}
\begin{center}
  \setlength{\tabcolsep}{0.2cm}
\begin{tabular}{|l  c   c   c l|}
\hline
$^A_\Lambda {\rm Z} (J^\pi,T)$ &   w/o YNN  & YNN\,($C'_{1}$,$C'_{3}$)   & YNN\,($C'_{1}$,$C'_{2}$,$C'_{3}$)  &  Exp.~\cite{HypernuclearDataBase}\\
 \hline
 \hline
{$^3_{\Lambda}\mathrm{H}(1/2^+,0)$} &  $0.121(4) $   & $0.125(4)$  & $0.155(3)$ & $0.164(43)  $  \\
\hline
$^4_{\Lambda}\mathrm{He}\,(0^+,1/2)$  &  $1.954(1)$   &   $ 2.027(3)$ &  $2.220(2)$ & $2.258(55) $  (average)   \\
&  &  &   & $2.169(42) $ $ (^4_{\Lambda}\mathrm{H})$ \\
&  &  &   & $2.347(36)$  $ (^4_{\Lambda}\mathrm{He})$ \\
\hline
$^4_{\Lambda}\mathrm{He}\,(1^+,1/2)$  &  $1.168(20)$   &   $ 1.010(11)$ & $0.984(12)$ &  $1.011(72) $  (average)  \\
&  &  &   & $1.081(46) $ $ (^4_{\Lambda}\mathrm{H})$ \\
&  &  &   &  $ 0.942(36)$ $ (^4_{\Lambda}\mathrm{He})  $\\
\hline
$^5_{\Lambda}\mathrm{He}(1/2^+,0)$  &  $ 3.518(20)$   &   $ 3.152(21)$ & $3.196(20)$ & $3.102(30)$\\
\hline
$^7_{\Lambda}\mathrm{Li}(1/2^+,0)$  & $5.719(56)$   &   $ 5.444(57)$ &  $5.623(52)$ & $5.619(60)$  \\
$^7_{\Lambda}\mathrm{Li}(3/2^+,0)$ & $5.522(70)$ &    $ 5.042(65)$ &  $5.040(57)$ & $4.927(60)$  \\
$^7_{\Lambda}\mathrm{Li}(5/2^+,0)$ & $3.440(66)$ &   $ 3.205(65)$ &  $3.356(60)$  & $3.568(60)$ \\
 \hline
  \end{tabular}
\end{center}
\caption{Separation energies for $A=3-7$ hypernuclei with angular momentum and parity $J^\pi$ and isospin $T$
in MeV, calculated without and with inclusion of
YNN three-body forces. See text for details on the
employed NN (YN) and 3N (YNN) potentials.
The number in parenthesis indicates the estimated extrapolation uncertainties.
}
\label{tab:BE}
\renewcommand{\arraystretch}{1.0}
\end{table*}

Note that by including decuplet baryons as degrees of freedom in chiral EFT, the pertinent parts of the
two-pion exchange $\La$NN and $\Si$NN 3BF are promoted to NLO as well as contributions that 
involve contact vertices. Therefore, we expect that
using decuplet saturation is a good starting point 
and should allow one to achieve a reliable semi-quantitative 
estimate for the effects of the $\La$NN and $\Si$NN 3BFs. 


{\bf Results.}--
Results for the separation energies of the
hypernuclei 
$^3_{\Lambda}\mathrm{H}$, $^4_{\Lambda}\mathrm{He}$,
$^5_{\Lambda}\mathrm{He}$, and $^7_{\Lambda}\mathrm{Li}$
without YNN force can be
found in the 2nd column of Tab.~\ref{tab:BE}.
We include in parenthesis the numerical uncertainty due to a 
necessary extrapolation in the model space size. 
Obviously, the separation energies based on the
two-body interaction alone are already fairly
close to the experimental values, cf. the
last column. 
Since, at present, no charge-symmetry breaking (CSB)
is included in the YN potential, we provide also
the average of the experimental $^4_\Lambda$H and 
$^4_\Lambda$He separation energies (for the 
$0^+$ and $1^+$ states) as benchmark for the comparison with the theoretical results. 

\begin{figure}[tbp] 
      \begin{center}          { \includegraphics[width=0.53\textwidth,trim={1.0cm 0.00cm 0.0cm 0 cm},clip]{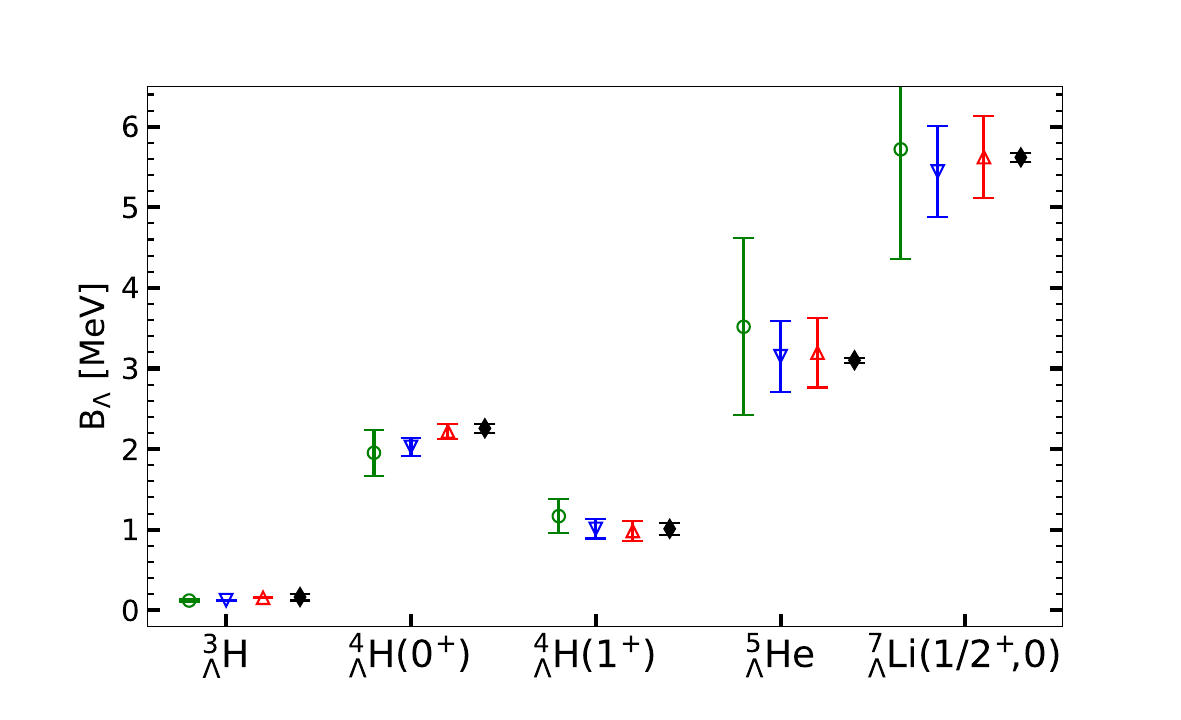}}
      \end{center}  
      \vskip -0.5cm
    \caption{Separation energies for $A=3-7$ 
    hypernuclei.
    Results are shown for the N$^2$LO YN potential
    without YNN force (green circles), and including YNN forces strictly (blue down-triangles) or partially constrained (red up-triangles) by decuplet saturation. The bars indicate the estimated truncation error of the chiral expansion, 
    see text. Experimental values (black diamonds)
    are taken from the chart of hypernuclides \cite{HypernuclearDataBase}.
    }
     \label{fig:Bayes}
         \end{figure} 
  
In a first step, we add $\La$NN and $\Si$NN 3BFs
based purely on decuplet saturation. In this case, there
are two independent LECs whose values are adjusted 
with the aim to reproduce  
the separation energies for the $0^+$ and
$1^+$ states of $^4_\Lambda$H/$^4_\Lambda$He
and  $^5_\Lambda$He. However, this can be only partially
achieved, see 3rd column of Tab.~\ref{tab:BE}.
While for the latter two cases the energies with
YNN force agree nicely with the experimental values within
their uncertainties, the result for the
$0^+$ state of $^4_\Lambda$H/$^4_\Lambda$He
remains noticeably below the empirical information. 
Regarding the other two hypernuclei, not considered when
adjusting the LECs of the 3BF, it turns out that the
result for $^3_\Lambda$H remains practically unchanged
while the separation energy for $^7_\Lambda$Li is now smaller than the experimental value.
The resulting LECs for the YNN force are  
$H_1 =-0.68\,F^{-2}_\pi$, $H_2=0.016\,F^{-2}_\pi$ where $F_{\pi} = 92.4 $~MeV and which 
is consistent with naturalness.  

Given the fact that with the decuplet saturation approximation alone it is not possible to simultaneously describe
the levels in the $A=4$ system and the separation energy of $^5_{\Lambda}\mathrm{He}$, 
in a second step, we go beyond that scenario. 
Specifically, we 
allow the LEC $C'_{2}$ to be different from zero while keeping $C'_{1}$ and $C'_{3}$ the same as in the initial fit. Note that
the $C'_2$ term introduces a 3BF that
depends explicitly on the spin of
the $\Lambda$ \cite{Hildenbrand:2024ypw},
cf. Eq.~(\ref{eq:LNNct}).
The corresponding separation energies are presented
in the 4th column of Tab.~\ref{tab:BE}, which have been achieved with a $C'_2=1310$~GeV$^{-5}$ of the expected size.
Obviously,
now all the $^4_{\Lambda}\mathrm{He/ H}(0^+,1^+)$ and $^5_{\Lambda}\mathrm{He}$ separation energies can be well reproduced
within the experimental uncertainty. Note that the spin dependent YNN force induces a modest shift of the separation energy of $^3_\Lambda$H which brings it in better agreement 
with the experimental average of the Mainz group
\cite{HypernuclearDataBase}. Also the $^7_\Lambda$Li separation energy is in remarkable agreement with experiment.  

To provide an estimate for the
theoretical uncertainty, we evaluated the truncation error of the chiral expansion based on the Bayesian approach \cite{Melendez:2017phj,Melendez:2019izc}. We learn the convergence of the chiral expansion based on the results that are computed using the LO, NLO and $\mathrm{N^2LO}$ YN interactions, see \cite{Le:2023bfj} for more details. At NLO and $\mathrm{N^2LO}$ we distinguish three cases, namely, without chiral YNN force, and including YNN forces with strict or with partial constraints from decuplet saturation. 
The separation energies for $A=3-7$ systems at $\mathrm{N^2LO}$  together with the theoretical uncertainties 
are displayed in  Fig.~\ref{fig:Bayes}. Expecting that the fits include the quantitatively most relevant parts of the YNN force, the chiral truncation errors are taken at $\mathrm{N^2LO}$ for the calculations including the chiral YNN forces, and at NLO otherwise. 
The experimental situation \cite{HypernuclearDataBase}, 
is indicated
by filled (black) diamonds. Obviously, the chiral truncation errors exceed the extrapolation uncertainties shown in the parentheses in Tab.~\ref{tab:BE}.
Note that the shifts due to the YNN forces are well within the 
NLO uncertainties in Fig.~\ref{fig:Bayes}. 
This indicates that the YNN contribution is
in line with the power counting.

\begin{figure}[tbp] 
  \begin{center}
  {\includegraphics[width=0.62\textwidth,trim={1.5cm 0.00cm 0.0cm 0.0cm},clip]{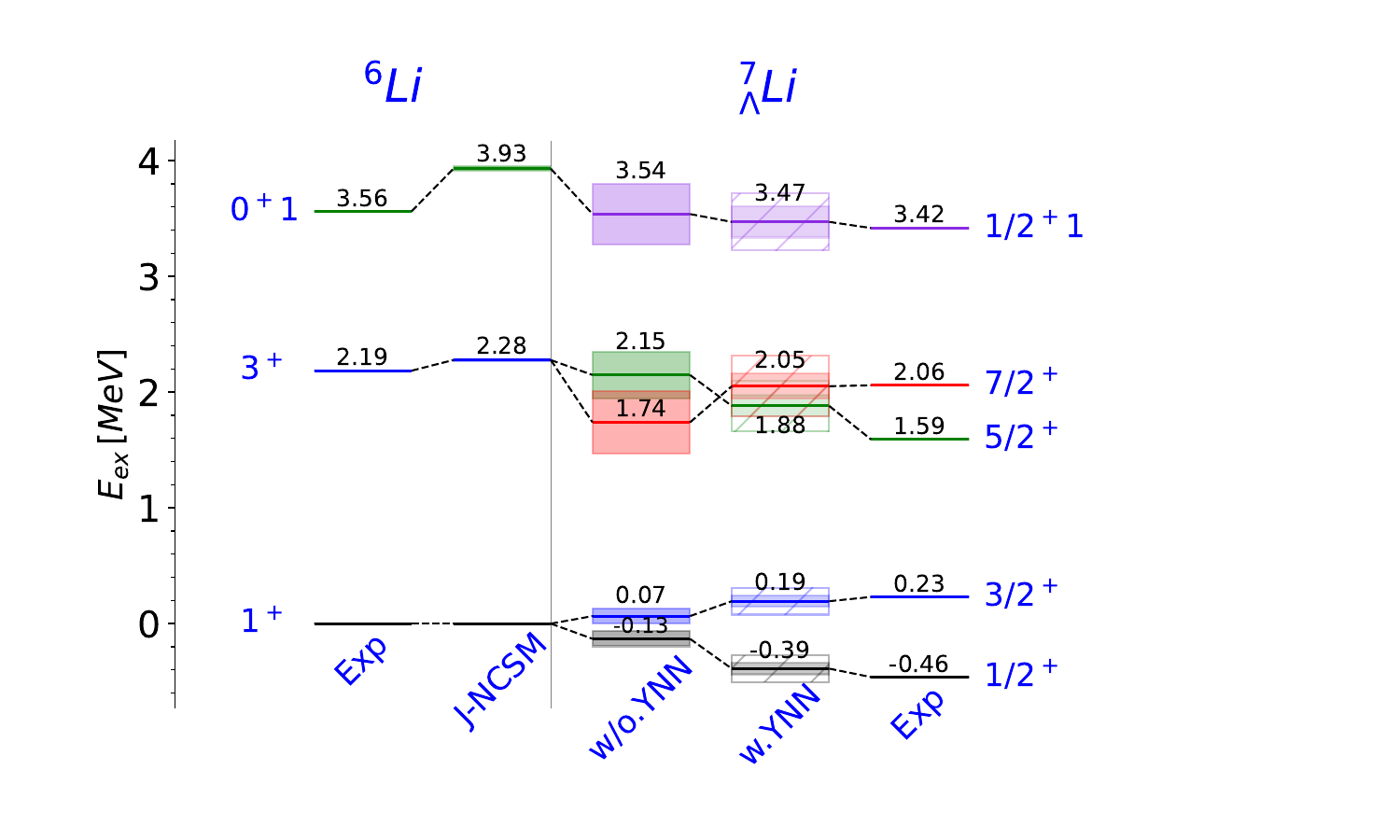}}
  \end{center}    
  \vskip -1.0cm
    \caption{Energy spectrum of $^7_{\Lambda}\mathrm{Li}$, calculated without and
    with YNN three-body forces. 
    Shaded areas show the extrapolated and chiral uncertainties at $\mathrm{NLO}$ and $\mathrm{N^2LO}$ for the case without and with chiral YNN forces, respectively. For the latter case, the hatched area represent uncertainties at $\mathrm{NLO}$. Experimental values are taken from \cite{Tamura:2012zz}. }
       \label{fig:7LiL}
     \end{figure} 

A detailed visualization of our predictions for the
energy spectrum of the $^7_{\Lambda}\mathrm{Li}$ hypernucleus
is provided in Fig.~\ref{fig:7LiL},
with the excitation spectrum of the core nucleus $^6\mathrm{Li}$ being shown on the right hand side. As one can see,  with the inclusion of the chiral 3NF, we reproduce the excitation energy of the $J^\pi=3^+$ state of $^6$Li fairly well. 
In order to further reduce the impact of the levels of the core nucleus,  we focus our discussion on the relative positions of the energy levels of 
$^7_\Lambda$Li and plot them with respect to the ground-state centroid energy \cite{Le:2019gjp}. The bands indicate the 
estimated uncertainties from the extrapolation and from the chiral convergence study at NLO  and  $\mathrm{N^2LO}$  for the cases without and with YNN forces, respectively. 
For the latter case, we also show the truncation error at NLO (hatched area). Without chiral YNN forces, we can only quantitatively reproduce the ground-state splitting and the isospin $T=1$ energy level. In particular, the level ordering of the $5/2^+$ and 
$7/2^+$ levels is opposite to the experiment. This behaviour is also observed for NLO13 and NLO19 versions of the chiral YN potentials, which differs from our earlier results reported in~\cite{Le:2019gjp} where the SRG-induced 3N and YNN forces were missing.
As clearly seen in the fourth column of Fig.~\ref{fig:7LiL}, the inclusion of the chiral YNN forces leads to significant improvement in the overall description of the $^7_{\Lambda}$Li spectrum. Specifically, the ground-state splitting as well as the $T=1$ level are consistent with experiment within the
estimated uncertainties, whereas the $5/2^+$ and $7/2^+$ levels are now correctly ordered. 
The shown results correspond to YNN forces with relaxed constraints from decuplet saturation,
i.e. involving all $C^{\prime}_{1-3}$ LECs. The results for the strict decuplet scenario are in between the cases w/o YNN and the full calculation and have been omitted to improve the readability of the figure. 

{\bf Summary.}--
We have performed a
study of light $\Lambda$ hypernuclei within chiral 
EFT up to N$^2$LO where, for the 
first time, chiral $\Lambda$NN and $\Sigma$NN three-body
forces are included consistently. The LECs
of the YNN forces were determined by fitting to the 
$^4_{\Lambda}\mathrm{H}$/$^4_{\Lambda}\mathrm{He}$ and
$^5_{\Lambda}\mathrm{He}$ separation energies, 
guided by constraints on their values from 
resonance saturation via decuplet baryons. 
Results for the hypernuclei $^3_{\Lambda}\mathrm{H}$, 
$^4_{\Lambda}\mathrm{H}$/$^4_{\Lambda}\mathrm{He}$,
$^5_{\Lambda}\mathrm{He}$, and $^7_{\Lambda}\mathrm{Li}$ could be achieved that agree well with experiments.  
Our calculations show that a consistent description
of $\Lambda$p scattering data and light hypernuclei
is possible, when including 3BFs in line with the
power counting of chiral EFT. 
One can certainly view that as confirmation of our present
knowledge of the $\Lambda$N-$\Sigma$N interaction
as represented by the chiral YN potentials. 
Nonetheless, one should not forget open issues
like the so far basically unconstrained interaction
in the higher partial waves of the $\Lambda$N
system \cite{Haidenbauer:2023qhf,Miwa:2022coz}
and certain indications that the $\Lambda$N interaction could be possibly overall slightly weaker~\cite{Mihaylov:2023ahn}. 
Finally, it is interesting to note that we observe a trend for repulsive effects of the required 3BFs 
with increasing $A$, 
which might be interpreted as support for one of the possible 
solutions of the hyperon puzzle in neutron stars.

\vskip 0.1cm
This project is part of the ERC Advanced Grant ``EXOTIC'' supported the European Research Council (ERC) under the European Union’s Horizon 2020 research and innovation programme (grant agreement No. 101018170). This work was further supported in part by the Deutsche Forschungsgemeinschaft (DFG, German Research Foundation) and the NSFC through the funds provided to the Sino-German Collaborative Research Center TRR110 ``Symmetries and the Emergence of Structure in QCD'' (DFG Project ID 196253076 - TRR 110, NSFC Grant No. 12070131001),
and by the MKW~NRW under the funding code NW21-024-A.
The work of UGM was supported in part by The Chinese Academy
of Sciences (CAS) President's International Fellowship Initiative (PIFI)
(grant no.~2025PD0022). We also acknowledge support of the THEIA net-working activity 
of the Strong 2020 Project. The numerical calculations were performed on JURECA
of the J\"ulich Supercomputing Centre, J\"ulich, Germany.

\bibliographystyle{unsrturl}

\bibliography{bibliography.bib}

\end{document}